# Pupil aberrations correction of the afocal telescope for the TianQin project

Zichao Fan[1], Zhengbo Zhu[2], Huiru Ji[1], Yan Mo[1], Hao Tan[1], Lujia Zhao[1], Shengyi Cao[2], AND Donglin Ma[2,3,*]

[1]*MOE Key Laboratory of Fundamental Physical Quantities Measurement & Hubei Key Laboratory of Gravitation and Quantum Physics, PGMF and School of Physics,*
*Huazhong University of Science and Technology, Wuhan 430074, China*
[2]*School of Optical and Electronic Information and Wuhan National Laboratory for Optoelectronics,*
*Huazhong University of Science and Technology, Wuhan 430074, China*
[3]*Shenzhen Huazhong University of Science and Technology, Shenzhen 518057, China*
*\*madonglin@hust.edu.cn*

**Abstract:** TianQin is a planned Chinese space-based gravitational wave (GW) observatory with a frequency band of $10^{-4}$ to 1Hz. Optical telescopes are essential for the delivery of the measurement beam to support a precise distance measurement between pairs of proof masses. As the design is driven by the interferometric displacement sensitivity requirements, the stability control of optical path length (OPL) is extremely important beyond the traditional requirement of diffraction-limited imaging quality. In a telescope system, the recurring tilt-to-length (TTL) coupling noise arises from the OPL variation due to the wavefront deformation and angular misalignment. The pupil aberrations are preferred option to understand the OPL specifications and further suppress TTL coupling noise. To correct the pupil aberrations, we derive primary pupil aberrations in a series expansion form, and then refine the formulation of merit function by combining the pupil aberration theory and traditional image aberration theory. The automatic correction of pupil aberrations is carried out by using the macro programming in the commercial optical software Zemax, leading to a high performance telescope design. The design results show that on one side the pupil aberrations have been corrected, and on the other side, its optical performance meets the requirements for TianQin project. The RMS wavefront error over the science field of view (FOV) is less than $\lambda/200$ and the maximum TTL coupling noise over the entire ±300 μrad FOV is 0.0034nm/μrad. We believe that our design approach can be a good guide for the space telescope design in any other space-based GW detection project, as well as other similar optical systems.

## 1. Introduction

The TianQin project is a space-borne gravitational waves (GWs) detector mission that operates in the millihertz frequency range [1-3]. The GWs observatory involves three spacecraft forming an equilateral triangle in a geocentric orbit in space, with arms approximately $10^5$ kilometers in length. Laser interferometry is used to monitor the distances between the test masses, where the signal to be obtained is the optical path length (OPL) between the transmitting transceiver system and the receiving transceiver system. The OPL signal from matching proof masses between different spacecrafts will correspond to the distortion of the equilateral triangle constellation caused by GWs, which is measured using well-known heterodyne interferometry techniques [4]. At the end of each arm, a space-borne telescope is equipped to expand and deliver the measurement beam to support precise distance measurements between pairs of proof masses. Since the afocal telescope serves for heterodyne interferometry detection, the design motivation and performance evaluation of the telescope is necessitated to differ from that of a traditional imaging system.

Afocal telescopes have been extensively studied in ongoing laser interferometer projects, such as LISA [5-10], DECIGO [11], Taiji [12,13], and TianQin [14,15]. The designs place great emphasis on addressing unusual requirements, including the scattered light, the wavefront quality, and OPL stability. The scattered light will introduce phase noise into the interferometric beat signal [16,17]. The expected tolerable level of scattered noises should be roughly in line with the received signal power from the far spacecraft. To avoid Narcissus reflection, an off-axis configuration is employed in almost all current designs, as the main source of scattered light on the detector comes from the backward scattering of the transmitted beam. Achieving ultra-high wavefront quality is of great necessity to support precision metrology, with the optical design specifications of the telescope exceeding the diffraction limit. Additionally, the OPL stability requirement for the telescope necessitates that the wavefront over the field should be smooth enough to prevent jitters from changing the OPL to the far spacecraft. After years of exploration, the telescopes described in these literatures are usually far beyond the diffraction limit. Nonetheless, there is still considerable potential for improving the design of diffraction-limited telescopes for space-based GW detection, particularly with regard to the tilt-to-length (TTL) coupling noise performance.

The angular misalignments between the reference beam and measuring beam can directly change the OPL traveled by the measurement beam, leading to the phenomenon of TTL coupling noise [18-19]. This coupling noise between the misalignments and the OPL readout is a significant noise source in precision interferometric length measurements. Therefore, a key goal in telescope design for space-based GW detection is to minimize or eliminate TTL coupling noise. Livas's group proposed that pupil aberrations could serve as a significant performance metric for the laser telescope in space GW detection and upgraded the design of LISA telescope by correcting pupil aberrations [10]. The so called pupil aberrations can be defined as the aberrations observed at the exit pupil when an object is placed at the entrance pupil for an optical system [20-22]. Given that the telescope system is afocal, pupil aberrations provide a more direct evaluation metric for the OPL stability between the entrance pupil and the exit pupil with field jittering, particularly when the entrance pupil is located at the test mass plane.

LISA team adopts the Seidel aberration theory to assess the afocal telescope's pupil aberration performance by interchanging the roles of the marginal and chief rays in the image aberration formula [10, 23-26]. Figure 1 illustrates the imaging principle from the point P on the entrance pupil to the point P' on the exit pupil, where the chief ray determines the imaging position and the marginal ray decides the imaging height [21]. The "pupil" for the conjugate system of both pupil planes is the image plane of normal imagery. After correcting for pupil aberrations, the improved LISA design 725-v7b satisfies all optical performance specifications, including wavefront error and TTL coupling noise. However, due to the lack of mature tools for analyzing pupil aberration performance, the systematic design method based on performance metrics regarding to pupil aberrations are not adequately developed.

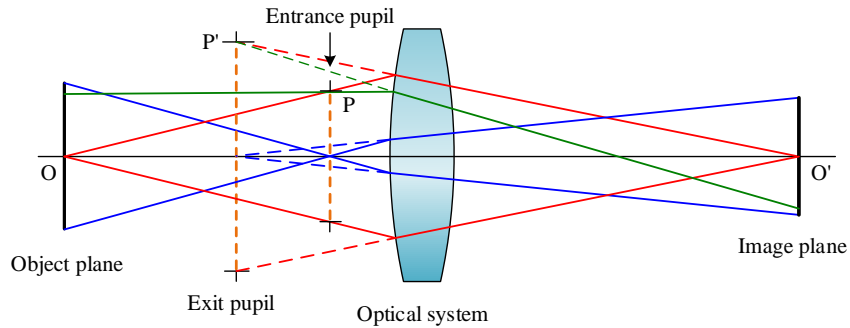

Fig.1 The function of the chief ray and marginal ray in pupil imaging

Our previous work about TianQin telescope design described wavefront aberration correction and related performance simulation, with no consideration on pupil aberrations [15]. The purpose of this paper is to present the recent development for the TianQin telescope design that specifically addresses the correction of pupil aberrations. An automated method based on the pupil aberration theory is developed to design an improved afocal catoptric telescope with the goal of minimizing the TTL noise. This study comprehensively discusses the optimization strategy and method for correcting the pupil aberrations, which differs from conventional imaging systems in ways such as the definition of the field of view (FOV), the construction of the merit function, and the performance metric for evaluation. We also develop tools for pupil aberration correction using the macro programming in the commercial optical software Zemax. The proposed design method is utilized to improve the TianQin telescope design, and its feasibility and validity are demonstrated and verified.

## 2. Design Specifications and special requirements

Table 1 lists the design specifications, which are derived from the requirements of TianQin project [14-15] and other LISA-like missions [5-7, 10, 12-13]. The working wavelength λ=1064nm is determined by the Nd:YAG laser used in the heterodyne interferometry Frequency and amplitude stabilization are the determining factors for this wavelength choice [22]. The design residual wavefront error (WFE) is allocated from the total budget of the entire system, with the Strehl ratio as the performance metric. Although the correction of pupil aberrations is taken as the main design goal, the optical system still needs to reach the diffraction limit for the correction of wavefront aberrations. If the WFE≤λ/20 in the single link, the corresponding total Strehl ratio value is above 0.8, which is generally considered the threshold for achieving the diffraction limit [27]. Large afocal magnification and the aperture are chosen to provide guarantees for efficient power transfer. The field of regard (FOR) is determined by the selected acquisition time, arm length, and orbit, and represents the FOV that the valid link can be established to acquire initial pointing signal [5]. The science FOV represents the angular range that the complete gravitational wave measurement can be carried out, considering the stability of orbit and pointing accuracy. The position of the entrance pupil is located in the test mass plane for the consideration of pupil aberrations. The TTL noise allocation derives from the OPL noise budget, and we expect that the differential change in OPL with angle can be as low as possible when the receiving beam passes through the afocal telescope.

**Table 1. Specifications of the TianQin telescope**

| Parameter | Specification |
|---|---|
| Wavelength | 1064nm |
| Wavefront quality (design residual) | ≤ λ/200 RMS@1064nm |
| Acquistion FOR | ±200μrad |
| Science FOV | ±7 μrad |
| Entrance pupil diameter | 220mm |
| Stop location | Test mass plane |
| TTL noise | ≤0.6nm/μrad |

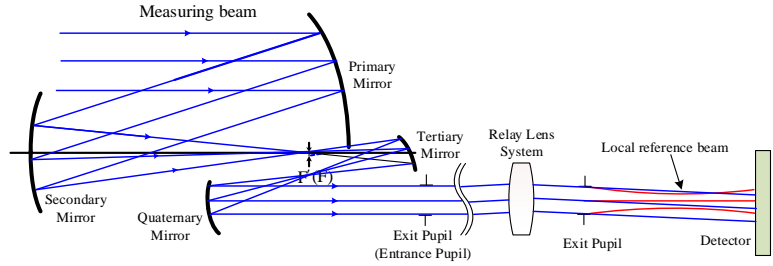

Fig.2 Simplified diagram shows the process where the measuring plane beam from another spacecraft is compressed to be a flat beam with aberrations caused by telescope, and then interfere with local reference Gaussian beam after passing the relay optical system.

Correcting pupil aberrations involves several considerations. Although this is a geometric optics concept, it is necessary to analyze the effect of phase aberrations on the interferometry in the TianQin mission. Figure 2 shows the telescope's role in light transfer and heterodyne interferometry [4, 18, 28-30]. Based on the far-field propagation's diffraction calculating results, the measuring beam received by the telescope can be simplified as a plane wave [31]. In the simplest case, we assume that the wavefront deformation of the measured beam during interferometry originates solely from the WFE of the telescope. Although the object and the image are in the spaces at infinity, a finite entrance pupil plane can be imaged to a finite exit pupil plane. The tilted measuring beam is rotated on the pupil plane instead of image plane. Therefore, pupil aberrations in the form of wavefront aberration is a more insightful merit function to optimize OPL in an afocal system.

On the other hand, when discussing system distortion, the transverse ray aberrations are usually chosen to evaluate the magnitude of aberrations. Figure 3 illustrates the pupil aberrations and reference Gaussian image points in an optical system. It is well-known that the image wavefront aberration, $W_i$, can be expanded as a power series [24]:

$$\begin{aligned}W_i =& b_1(x_p^2 + y_p^2) + b_2 yy_p + b_3 y^2 + c_1(x_p^2 + y_p^2)^2 + c_2 yy_p(x_p^2 + y_p^2) + c_3 y^2 y_p^2 \\&+ c_4 y^2(x_p^2 + y_p^2) + c_5 y^3 y_p + c_6 y^4 + d_1(x_p^2 + y_p^2)^3 + d_2 yy_p(x_p^2 + y_p^2)^2 + \\&d_3 y^2 y_p^2(x_p^2 + y_p^2) + d_4 y^2(x_p^2 + y_p^2)^2 + d_5 y^3 y_p(x_p^2 + y_p^2) + d_6 y_p^3 + d_7 y^4 y_p^2 \\&+ d_8 y^4(x_p^2 + y_p^2) + d_9 y^5 y_p + d_{10} y^6 + \text{terms with higher order.}\end{aligned} \quad (1)$$

The primary aberrations in an optical system are commonly represented by the five three-order terms with coefficients $c_1$ to $c_5$. The terms that only depends on the field, such as $c_6$, $d_{10}$…, are typically assumed to be zero, based on the assumption that wavefront aberrations should vanish in the center of the pupil [20]. Since this term is independent of pupil coordinates, it is typically ignored when differentiating the wavefront aberration with respect to $x_p$ and $y_p$ to derive the transverse ray aberrations. Nevertheless, the aberrations that only depends on the field can cause the OPL to change as the field coordinate of the characteristic rays, which contradicts our design goals [32].

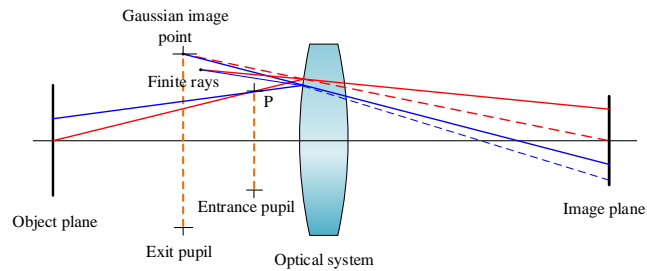

Fig.3 Geometric spot for pupil imaging. The aberrations are exaggerated for clarity.

Therefore, it makes sense to consider realizing pupil aberration correction to improve the design. In the following sections, we will introduce the study about correcting the primary pupil aberrations at the third-order level, and subsequently optimize the optical system variables and evaluate pupil aberrations with higher-order terms.

## 3. Primary Pupil Aberration

We wish to clarify the physical meaning of coordinate transformation of pupil aberration in the derivation under the geometric optical framework. The total aberrations can be computed as the sum of aberration contributions from various surfaces of the optical system.

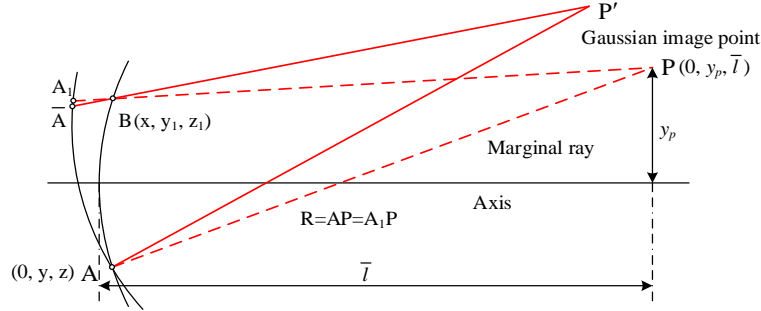

Fig.3 Diagram of OPL in a pupil imaging system

Welford derived the optical path difference introduced at a single refracting surface, which is also applicable to pupil imaging [24]. As shown in Fig. 3, PA and PA$_1$ are the radius of the reference sphere with the center at gaussian image point P. P' is the real image when the pupil aberrations are present. PA$_1$ intersects the refracting surface at point B $(0, y_1, z_1)$, and the extension of P'B intersects the reference surface at point $\overline{A}$. The total wavefront aberration can be written as $W=\sum\Delta\{n(\overline{A}B)\}$ and we have $\overline{A}B = A_1B$ when the higher order aberrations are negligible in deriving the primary aberration [24-25]. The $\Sigma$ notation indicates the summation operator taken over all the refracting surfaces, and the $\Delta$ signifies the incremental change in refraction. Then based on a simple geometric relationship, we obtain

$$W=\sum\Delta\{n(\mathrm{PA}-\mathrm{PB})\}, \qquad (2)$$

where PA and PB can be expressed as:

$$\mathrm{PA}=\sqrt{(y_p - y)^2 + (\overline{l} - z)^2} \qquad (3)$$

$$\mathrm{PB}=\sqrt{x^2 + (y_p - y_1)^2 + (\overline{l} - z_1)^2}, \qquad (4)$$

where the $\overline{l}$ is the gaussian conjugate distance and $y_p$ is the gaussian image height in pupil imaging. Eq. (4) can be expanded by the Taylor series, and we neglect the terms of order higher than four-order:

$$\mathrm{PB}=(\overline{l}-z_1)+\frac{1}{2}\cdot(\overline{l}-z_1)\cdot\frac{x^2+(y_p-y_1)^2}{(\overline{l}-z_1)^2}-\frac{1}{8}\cdot\left\{\frac{x^2+(y_p-y_1)^2}{(\overline{l}-z_1)^2}\right\}^2, \qquad (5)$$

where $z_1$ represents the sag of refracting surface. Let the radius of curvature of the surface be $r$, then according to geometric relations we obtain:

$$\left(x^2+y_1^2\right)+(r-z_1)^2=r^2. \qquad (6)$$

Solving the roots of the quadratic Eq. (6) and expanding it with the Taylor series we get:

$$z_1 = r - r\cdot(1-\frac{(x^2+y_1^2)}{2r^2}-\frac{(x^2+y_1^2)^2}{8r^4}) = \frac{(x^2+y_1^2)}{2r}+\frac{(x^2+y_1^2)^2}{8r^3}. \qquad (7)$$

Substituting the Eq. (7) to Eq. (5), we have:

$$\mathrm{PB} = \bar{l} - \frac{1}{2r}(x^2 + y_1^2) + \frac{1}{2\bar{l}}\{x^2+(y_p - y_1)^2\} - \frac{1}{8r^3}(x^2 + y_1^2)^2 +$$
$$\frac{1}{4r\bar{l}^2} \cdot (x^2 + y_1^2)\{x^2+(y_p - y_1)^2\} - \frac{1}{8\bar{l}^3}\{x^2+(y_p - y_1)^2\}^2. \tag{8}$$

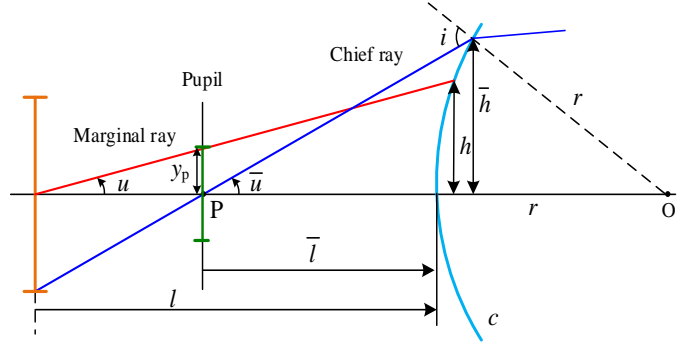

Fig.4 The derivation of paraxial invariants in a pupil imaging system

By simple transformation using Lagrange invariants, $y_p$ in Eq. (8) can be substituted by paraxial quantities. Figure 4 depicts the configuration of pupil imaging and shows two characteristics rays along with some basic paraxial quantities. From this figure, we can infer that $y_p = (\bar{l} - l)u$, and we can further eliminate the aperture angle $u$ from the expression and obtain:

$$y_p = \bar{l}h \cdot (\frac{1}{\bar{l}} - \frac{1}{l}) \tag{9}$$

It can be seen that the paraxial angle of incidence $i = \bar{h}c + \bar{u}$, and from the Snell's law we can obtain:

$$n'(\bar{h}c + u') = n(\bar{h}c + u). \tag{10}$$

Each side of Eq. (10) is denoted by $\bar{A}$, which is called the refraction invariant. Similarly, we can define another refraction invariant $A$ for the marginal ray. Then we can use these invariants to define the paraxial quantity $h_o = y_1 - y$. This eliminates the $y_1$ from the equation:

$$y_1 = h_0 + y \quad ; \quad y_1^2 = h_0^2 + 2h_0 y_1 + y_1^2 \tag{11}$$

Substituting the expressions in Eq. (9) and Eq. (11) into Eq. (8) we can get:

$$PB = B_1(x^2 + h_0^2)^2 + B_2(x^2 + h_0^2)h_0 y + B_3 y^2 h_0^2$$
$$+ B_4 y^2 (x^2 + h_0^2) + B_5 y^3 h_0 + B_6 y^4$$

$$B_1 = -\frac{1}{8}(\frac{1}{r^3} - \frac{2}{r\bar{l}^2} + \frac{1}{\bar{l}^3}) \qquad B_4 = -\frac{1}{4}(-\frac{1}{rl^2} - \frac{1}{r\bar{l}^2} + \frac{1}{l^2\bar{l}} + \frac{1}{r^3}) \tag{12}$$

$$B_2 = -\frac{1}{2}(\frac{1}{r^3} + \frac{1}{l\bar{l}^2} - \frac{1}{rl\bar{l}} - \frac{1}{r\bar{l}^2}) \quad B_5 = -\frac{1}{2}(\frac{1}{l^3} - \frac{1}{l^2} - \frac{1}{2rl\bar{l}} + \frac{1}{2r^3})$$

$$B_3 = -\frac{1}{2}(\frac{1}{l^2\bar{l}} - \frac{2}{rl\bar{l}} + \frac{1}{r^3}) \qquad B_6 = -\frac{1}{8}(\frac{1}{r^3} - \frac{2}{rl^2} - \frac{l}{l^4})$$

If we let $x=0$ and $y_1=y$ in Eq. (12), PA can be quickly acquired as:

$$PA = B_6 y^4 \tag{13}$$

Now, by subtracting the Eq. (12) from Eq. (13), we can obtain the expression for PB−PA in Eq. (2).

We will only derive the term with coefficient $B_1$ as an example. However, it is worth noting that the derivation process for the other terms can be completed in a similar way. Further simplification involves dealing with the Δ symbol, which indicates that we need look for

quantities that remain constant before and after refraction. The well-known Gaussian imaging formula is:

$$\frac{n'}{l'} - \frac{n}{l} = (n'-n)c. \tag{14}$$

Rewrite Eq. (14) in the following form:

$$n(c-\frac{1}{l}) = n'(c-\frac{1}{l'}). \tag{15}$$

We can find that the quantity $n(c-1/l)$ is an invariant on refraction. Then, the first term of the primary pupil aberration expression can be transformed into:

$$-\frac{1}{8}\sum_{All\ surface}\Delta\left\{\left\{c^2 n(c-\frac{1}{l})\right\}+\Delta\left\{\frac{1}{n\bar{l}}\left[n(c-\frac{1}{l})\right]^2\right\}\right\}(x^2+h_0^2)^2$$

$$= -\frac{1}{8}\sum_{All\ surface}\Delta\left\{\frac{1}{n\bar{l}}\left[n(c-\frac{1}{l})\right]^2\right\}(x^2+h_0^2)^2. \tag{16}$$

To define the aberration coefficients, we use the full aperture and the full field. The aberrations at other positions can be obtained by proportional relations. In the meridian section where $x=0$, we can obtain the value of $h_0$ by tracing a paraxial ray from an axial object point with the conjugate distance $\bar{l}$. We let the $h_0 = \bar{y}$, which corresponds paraxially to $\bar{h}$. With this, we can simplify the Eq. (16) to the familiar form:

$$= -\frac{1}{8}\sum_{All\ surface}\bar{y}\Delta\left\{\frac{1}{n}\cdot(\frac{\bar{y}}{\bar{l}})\left[n\bar{y}(c-\frac{1}{l})\right]^2\right\}$$

$$= -\frac{1}{8}\sum_{All\ surface}\bar{A}^2 \bar{y}\Delta(\frac{\bar{u}}{n}) = -\frac{1}{8}\bar{S}_I \tag{17}$$

The coefficients of other primary pupil aberrations can be expressed as:

$$\begin{cases}
\sum_{All\ surface}\Delta(nB_2)\cdot(x^2+h_0^2)h_0 y = -\frac{1}{2}\sum_{All\ surface}A\bar{A}\bar{y}\Delta(\frac{\bar{u}}{n}) = -\frac{1}{2}\bar{S}_{II} \\
\sum_{All\ surface}\Delta(nB_3)y^2 h_0^2 = -\frac{1}{2}\sum_{All\ surface}A^2 \bar{y}\Delta(\frac{\bar{u}}{n}) = -\frac{1}{2}\bar{S}_{III} \\
\sum_{All\ surface}\Delta(nB_4)y^2(x^2+h_0^2) = -\frac{1}{4}\sum_{All\ surface}A^2 \bar{y}\Delta(\frac{\bar{u}}{n})+\bar{H}^2 c\Delta(\frac{1}{n}) = -\frac{1}{4}\bar{S}_{III}-\frac{1}{4}\bar{S}_{IV} \\
\sum_{All\ surface}\Delta(nB_5)y^3 h_0^2 = -\frac{1}{2}\sum_{All\ surface}\left\{\frac{A^3}{\bar{A}}\bar{y}\Delta(\frac{\bar{u}}{n})+\frac{A}{\bar{A}}\bar{H}^2 c\Delta(\frac{1}{n})\right\} = -\frac{1}{2}\bar{S}_V
\end{cases}, \tag{18}$$

where $\bar{H}$ is defined as $\bar{H} = n\bar{u}y_p = nu\bar{u}(\bar{l}-l)$, which is the Smith-Helmholtz invariant in the pupil imagery system. It is obviously $\bar{H} = -H$ in the same optical system.

In a polar coordinate, the wavefront aberration can be expressed in term of these primary pupil aberration sums with the normalized field coordinate $\rho$ and pupil coordinate $\eta$:

$$W(\rho,\varphi,\eta) = \frac{1}{8}\bar{S}_I\eta^4 + \frac{1}{2}\bar{S}_{II}\rho\eta^3\cos\varphi + \frac{1}{2}\bar{S}_{III}\rho^2\eta^2\cos\varphi +$$

$$\frac{1}{4}(\bar{S}_{III}+\bar{S}_{IV})\rho^2\eta^2 + \frac{1}{2}\bar{S}_V\rho\eta^3\cos\varphi. \tag{19}$$

From Eq. (19), the effects of phase aberrations can be quantitatively discussed on the basis of geometrical optics in the design.

## 4. Coaxial System Design

The afocal telescope design for the TianQin mission currently adopts a near diffraction-limited off-axis four-mirror configuration. During the coaxial design phase, the quaternary mirror (QM) contributes very little to the overall optical power of the system, and can even be a flat mirror without optical power. Therefore, an initial Cassegrain-style three-mirror configuration with a relay image can adequately satisfy the design requirements [33-36]. Figure 5 shows a ray tracing of an on-axis Cassegrain design for the TianQin mission, including a real relay image. In the sequence of ray propagation, the system consists of primary mirror (PM), secondary mirror (SM), and tertiary mirror (TM). The aperture stop of system is behind the PM, where the test mass is situated, and all mirrors are designed as spherical surfaces.

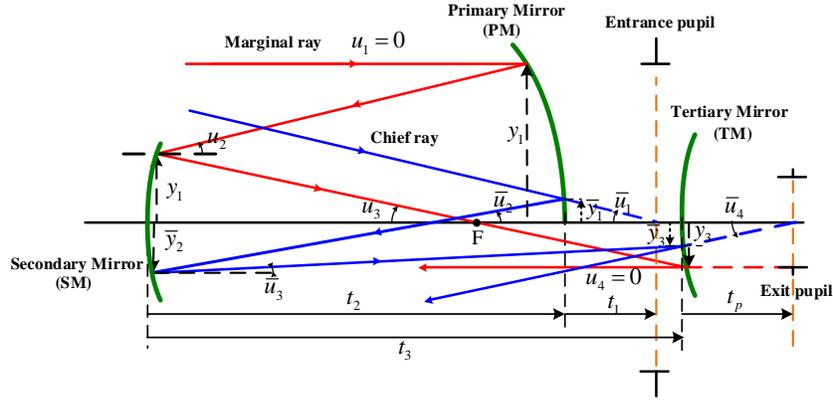

Fig.5 Ray tracing of initial coaxial structure

The curvature radius of the PM, SM and TM are respectively $R_1$, $R_2$, and $R_3$. $t_1$, $t_2$, $t_3$ and $t_4$ represent the distance between the entrance pupil and PM, the distance between PM and the SM, the distance between SM and the TM, and the distance between TM and the image plane respectively. The transmission media refractive index is set as $n_1 = n_2' = n_3 = 1$, $n_1' = n_2 = n_3' = -1$. The height $y$ of the intersection point of the ray with the surface and the paraxial angle $u$ of the ray's inclination can be calculated according to paraxial $y$-$u$ method:

$$\begin{cases} n_i' u_i' = n_i u_i - y_i \varphi_i \\ y_{i+1} = y_i + u_i' t_i' \\ t_i' = t_{i+1} \\ u_i' = u_{i+1} \end{cases}, \quad (20)$$

where $\varphi_i$ represents the optical power of the $i$-th mirror, which can be determined by:

$$\varphi_i = \frac{(n_i' - n_i)}{R_i}. \quad (21)$$

From Eqs. (17), (18), (20) and (21), the primary pupil aberration coefficients can be expressed as the functions of $y_1$, $\bar{y}_1$, $\varphi_i$ (i =1, 2, 3), and $t_i$ (i =1, 2, 3, 4) as Eqs. (22)-(26) [36].

$$\begin{aligned}\bar{S}_I = \frac{1}{4t_1^4} (\varphi_1 t_1^2 \bar{y}_1^4 (2 + \varphi_1 t_1)^2 + \varphi_2 (t_1 + t_2 + \varphi_1 t_1 t_2)^2 (2 + \varphi_2 (t_1 + t_2) + \varphi_1 t_1 \bar{y}_1^4 (2 + \varphi_2 t_2))^2 + \\ \varphi_3 (t_1 \bar{y}_1 - t_3 \bar{y}_1 (1 + \varphi_2 (t_1 + t_2) + \varphi_1 (t_1 + \varphi_2 t_1 t_2)) + t_2 (\bar{y}_1 + \varphi_1 t_1 \bar{y}_1))^2 (2 \bar{y}_1 + 2 \varphi_1 t_1 \bar{y}_1 + \\ 2 \varphi_2 \bar{y}_1 (t_1 + t_2 + \varphi_1 t_1 t_2) + \varphi_3 (t_1 \bar{y}_1 - (1 + \varphi_2 (t_1 + t_2) + \varphi_1 t_3 \bar{y}_1 (t_1 + \varphi_2 t_1 t_2)) + t_2 (\bar{y}_1 + \varphi_{11} t_1 \bar{y}_1)))^2)\end{aligned} \quad (22)$$

$$\bar{S}_{II} = \frac{y_1}{4t_1^3}(\varphi_1^2 t_1^2 \bar{y}_1^3(2+\varphi_1 t_1) + \varphi_2(t_1+t_2+\varphi_1 t_1 t_2)^2(\varphi_2+\varphi_1(2+\varphi_2 t_2))(2+\varphi_2(t_1+t_2) +$$
$$\varphi_1 t_1 \bar{y}_1^3(2+\varphi_2 t_2)) - 2\varphi_3 t_1(-1+\varphi_2 t_3 + \varphi_1(t_3+t_2(-1+\varphi_2 t_3)))(\varphi_1 y_1 - 1/2\varphi_3 y_1(-1+\varphi_2 t_3 +$$
$$\varphi_1(t_3+t_2(-1+\varphi_2 t_3))) + \varphi_2(y_1+\varphi_1 t_2 y_1))(t_1 \bar{y}_1 - t_3 \bar{y}_1(1+\varphi_2(t_1+t_2) + \varphi_1(t_1+\varphi_2 t_1 t_2)) + \quad (23)$$
$$t_2(\bar{y}_1+\varphi_1 t_1 \bar{y}_1))(2\bar{y}_1 + 2\varphi_1 t_1 \bar{y}_1 + 2\varphi_2 \bar{y}_1(t_1+t_2+\varphi_1 t_1 t_2) + \varphi_3(t_1 \bar{y}_1 - t_3 \bar{y}_1(1+\varphi_2(t_1+t_2) +$$
$$\varphi_1(t_1+\varphi_2 t_1 t_2)) + t_2(\bar{y}_1+\varphi_1 t_1 \bar{y}_1))))$$

$$\bar{S}_{III} = \frac{1}{4t_1^4}(\varphi_1 t_1^2(2+\varphi_1 t_1)^2 \bar{y}_1^4 + \varphi_2 \bar{y}_1^4(t_1+t_2+\varphi_1 t_1 t_2)^2(2+\varphi_2(t_1+t_2) + \varphi_1 t_1(2+\varphi_2 t_2))^2$$
$$-\varphi_3 t_1 y_1(-1+\varphi_2 t_3 + \varphi_1(t_3+t_2(-1+\varphi_2 t_3)))(t_1 \bar{y}_1 - t_3 \bar{y}_1(1+\varphi_2(t_1+t_2) + \varphi_1(t_1+\varphi_2 t_1 t_2)) + \quad (24)$$
$$t_2(\bar{y}_1+\varphi_1 t_1 \bar{y}_1))(2\bar{y}_1 + 2\varphi_1 t_1 \bar{y}_1 + 2\varphi_2 \bar{y}_1(t_1+t_2+\varphi_1 t_1 t_2) + \varphi_3(t_1 \bar{y}_1 - t_3 \bar{y}_1(1+\varphi_2(t_1+t_2) +$$
$$\varphi_1(t_1+\varphi_2 t_1 t_2)) + t_2(\bar{y}_1+\varphi_1 t_1 \bar{y}_1)))^2)$$

$$\bar{S}_{IV} = \frac{y_1^2 \bar{y}_1^2(\varphi_1+\varphi_2+\varphi_3)}{t_1^3} \quad (25)$$

$$\bar{S}_V = \frac{y_1^3 \bar{y}_1}{4}(\frac{\varphi_1^4 t_1}{2+\varphi_1 t_1} + (\frac{\varphi_2(t_1+t_2+\varphi_1 t_1 t_2)^2(\varphi_2+\varphi_1(2+\varphi_2 t_2))^3)}{t_1(2+\varphi_2(t_1+t_2)+\varphi_1 t_1(2+\varphi_2 t_2))} + (\varphi_3(-\varphi_3+\varphi_2(-2+\varphi_3 t_3) +$$
$$\varphi_1(-2-\varphi_3 t_2+\varphi_3 t_3+\varphi_2 t_2(-2+\varphi_3 t_3)))^3(t_3+t_2(-1+\varphi_2 t_3)+t_1(-1+\varphi_2 t_3+\varphi_1(-t_2+t_3$$
$$+\varphi_2 t_2 t_3)))^2)/(t_1(-2-\varphi_3 t_1-\varphi_3 t_2+\varphi_3 t_3+\varphi_2(t_1+t_2) (-2+\varphi_3 t_3)+\varphi_1 t_1(-2-\varphi_3 t_2+\varphi_3 t_3$$
$$+\varphi_2 t_2(-2+f3\,t3)))) - \frac{2\varphi_1^3 \bar{y}_1}{2t_1+\varphi_1 t_1^2} + \frac{2\varphi_2^2 \bar{y}_1(t_1+t_2+\varphi_1 t_1 t_2)(\varphi_2+\varphi_1(2+\varphi_2 t_2))}{t_1^2(2+\varphi_2(t_1+t_2)+\varphi_1 t_1(2+\varphi_2 t_2))} + \quad (26)$$
$$+(2\varphi_3^2(\varphi_3+\varphi_2(-2+\varphi_3 t_3)+\varphi_1(-2-\varphi_3 t_2+\varphi_3 t_3+\varphi_2 t_2(-2+\varphi_3 t_3)))(t_3+t_2(-1+\varphi_2 t_3)+$$
$$t_1(-1+\varphi_2 t_3+\varphi_1(-t_2+t_3+\varphi_2 t_2 t_3)))\bar{y}_1)/(t_1^2(-2-\varphi_3 t_1-\varphi_3 t_2+\varphi_3 t_3+\varphi_2(t_1+t_2)$$
$$(-2+\varphi_3 t_3) + \varphi_1 t_1(-2-\varphi_3 t_2+\varphi_3 t_3+\varphi_2 t_2(-2+\varphi_3 t_3)))))$$

To ensure the optimal performance of the afocal telescope for the TianQin mission, several constraints have been applied to the overall layout of the optical system. The aperture of the primary mirror has been set as 220mm, which not only satisfies the requirements of energy transmission but also ensures convenience in manufacturing and testing. Additionally, the optical length of the telescope has been limited to 400mm, excluding the supporting structure. To initiate the design process, we set the initial values of $y_1 = 110$ mm, $\bar{y}_1 = 5$ mm, $t_1 = -500$ mm and $t_3 = 250$mm. Further design and optimization will be carried out using the ZEMAX [37] commercial software. Due to the lack of pupil aberration calculation programs in current commercial software, the initial structural parameters of the coaxial system are obtained based on the Zemax Programming Language (ZPL) macro described in Table 2. At this phase, the QM is a flat mirror without any optical power. The initial design prescriptions, including the curvature radii of three mirrors and the distances between them, are listed in Table 3.

**Table 2. Programming description for the optimization based on primary pupil aberrations**

| ZPL macro 1: Pupil aberrations correction in a coaxial system |
|---|
| Input: $y_1$, $\bar{y}_1$, $t_1$, $t_3$ |
| Output 1: Primary pupil aberrations coefficient for wavefront power series expansion |
| Output 2: Primary pupil aberration value for merit function |
| 1:    Field Sampling Module |
| 2:    For m=1 to Nm // determines how many radial arms of rays in the circularly pupil |
| 3:      for n=1 to Nn do // determines how many rays are traced at each field |

```
4:      Fov(m, 2*n-1) = (m-1)*2*pi/Nm
5:      Fov(m, 2*n) = (n-1)*Maxfield/(Nn-1)
6:     end for
7:    end for
8:    Ray Tracing Module
9:    For t=1 to nfield do // nfield = Nm in rotationally symmetric system
10:    hy(1, t)=Fov(t,2)/Maxfiled // Calculating the normalized field coordinates
11:    RAYTRACE hx, hy(1,t), 0, 0, PWAV //marginal ray
12:    Record the height Hm(i,t) and angle Im(i,t) // i is serial number of current surface
13:    RAYTRACE hx, hy(1,t), 0, 0, PWAV //chief ray
14:    Record the height Hc(i,t) and angle Ic(I,t)
15:    End for
16:    For p=1 to Nm
17:    For q=1 to Nn do
18:    Calculate the $\bar{S}_I$, $\bar{S}_{II}$, $\bar{S}_{III}$, $\bar{S}_{IV}$, and $\bar{S}_V$ using Eqs. (22)-(26)
19:    Calculating the W(p,q) using Eq. (19)
20:    End for
21:    End for
22:    $W=\sum w_{pq} \cdot W(p,q)$
Optimization algorithm: damped least square method
```

**Table 3. Coaxial initial structure parameters**

| Surface | Radius(mm) | Distance(mm) | Conic |
|---|---|---|---|
| Stop | - | -500 | - |
| PM | -644.453 | -300 | -0.916 |
| SM | -48.596 | 350 | - |
| TM | -189.028 | -60 | - |
| QM | Infinity | - | - |
| Image | - | - | - |

## 5. Off-axial System Design and Optimization

The coaxial afocal telescope is transformed to an off-axis system with conic surfaces. TTL coupling noise generated by geometrical wavefront error can be suppressed efficiently by making the OPL uniform or at least symmetric over the FOV. The WFE should be kept as uniform as possible and have a uniform distribution at the sampled FOVs in all directions. The imaging quality of TianQin telescope is evaluated by both imaging RMS WFE and pupil aberrations in the optimization process.

However, due to the limitations of existing programs, ZPL macros are still used as an auxiliary tool for subsequent optimization. Thompson's nodal aberration theory suggests that the misalignment of the system only affects the aberration nodal position. [38-40]. The ZPL macro described in Table 2 is used to calculate the wavefront aberration coefficient in the coaxial system, which partially characterizes the pupil aberration level [41-42]. Furthermore, the uncorrected pupil aberrations can be reflected in the locations where the chief ray crosses the pupil plane. In an ideal optical system without any aberration, the chief rays tracing from any field should pass through the center of the exit pupil plane. Hence, we use the RMS radius of the chief ray on the pupil plane to quantitatively assess the level of pupil aberration [15]:

$$R_{CRMS}=\sum_{i=1}^{N}\frac{\sqrt{(x_i-\bar{x})^2+(y_i-\bar{y})^2}}{N}, \qquad (27)$$

where $N$ is the number of sampling points, $x_i$ is the $x$ coordinate of $i$-th sampling point on the pupil plane, $y_i$ is the $y$ coordinate of $i$-th sampling point on the pupil plane, the notation with bar is the corresponding arithmetic mean. $R_{CRMS}$ can be a suitable metric when the optimization involves high order aberrations. In our previous design, no pupil aberration correction was implemented, and the results showed that $R_{CRMS}$=0.729 mm at the FOV= 220 µrad.

Table 3. Programming description for the optimization in the off-axial design

| ZPL macro 2: Pupil aberrations correction in an off-axial system |
|---|
| Input 1: Exit pupil position // Called from the Zemax main program |
| Output 1: Coordinates on exit pupil plane // Called from ZPL macro 1 |
| Output 2: $R_{CRMS}$ |
| 1:    For m=1 to Nm |
| 2:     For n=1 to Nn do |
| 3:     // Field Sampling |
| 3:     Fov(m,2*n-1) = (m-1)*2*pi/Nm |
| 4:     Fov(m,2*n) = (n-1)*Maxfield/(Nn-1) |
| 5:     hx(m,n)= cos(Fov(m,2*n-1))*Fov(m,2*n)/Maxfiled |
| 6:     hy(m,n)= sin(Fov(m,2*n-1))*Fov(m,2*n)/Maxfiled |
| 7:     // Ray Tracing |
| 8:     RAYTRACE hx(m,n), hy(m,n), 0, 0, PWAV |
| 9:     Record the x-coordinate on exit pupil plane Px(m,n) |
| 10:    Record the y-coordinate on exit pupil plane Py(m,n) |
| 10:    End for |
| 11:    End for |
| 12:    $\bar{x} = \dfrac{\sum\sum Px(m,n)}{Nm*Nn}$ ;  $\bar{y} = \dfrac{\sum\sum Py(m,n)}{Nm*Nn}$ |
| 13:    Calculating the $R_{CRMS}$ using Eq. (23) |
| 14:    Plot the "chief ray spot diagram"// in MATLAB software |
| Optimization algorithm: damped least square method |

To meet the demanding requirements of correcting pupil and image plane aberrations simultaneously, a progressive strategy is employed during the design process, gradually increasing the number of aspheric terms. Lens parameters, including the radius of curvature, aspherical coefficients of the TM and QM, decenter value in Y-direction, and tilt about the x-axis, are set as variables in the optimization. The merit function comprises the default wavefront errors generated by Zemax, boundary conditions, and ZPL macros described in Tables 2 and 3. A more detailed description of optimization in software is beyond the scope of this article, and can be found in our previous work. After repeated optimization, the design of the TianQin telescope with pupil aberration correction is obtained.

## 6. Design results and performance

The layout of the final off-axial design, along with rays for 3 fields (-200 µrad, 0, +200 µrad measured in receiver-mode object space) is illustrated in Fig. 4. The PM is a large-diameter paraboloid mirror, while the other three mirrors are even-order aspheric surfaces. An intermediate image plane is placed inside the system so that the stray light outside the FOV can be shielded by the field stop placed at the focal plane. In the coordinate system shown in this figure, the real exit pupil is located 98.77mm behind the vertex of PM along the z-axis. The combination of the TM and QM directs the laser beam to the optical bench, where the beam measures about 1.5mm in width. The fundamental structural parameters and even-order aspheric coefficients are detailed in Table 4 and Table 5.

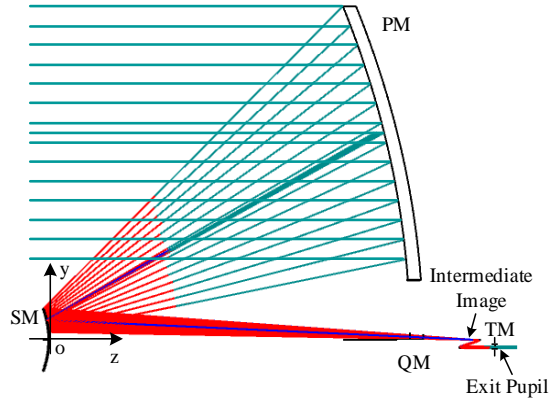

Fig.6 Optical layout of the TianQin telescope design

**Table 4. Parameters of final off-axial structures**

| Surface | Radius (mm) | Distance (mm) | Conic | Decenter Y (mm) | Tilt about x (°) | Size (mm) |
|---|---|---|---|---|---|---|
| Stop | - | -500 | - | - | - | - |
| PM | -708.325 | -320 | -0.916 | -180 | - | 230 |
| SM | -75.103 | 382.363 | -5.409 | - | - | 55 |
| TM | -195.349 | -18.68 | 16.46 | 0.125 | -8.1 | 2.5 |
| QM | 49.017 | - | 18.494 | 7.69 | -5.5 | 9.1 |
| Image | - | - | - | - | - | - |

The wavefront errors for each field are presented in Fig. 7, revealing excellent RMS wavefront errors and wavefront consistency. Figure 8 depicts the locations where the chief rays intersect the exit pupil plane for a set of fields. The blue circles represent the chief rays in the scientific FOV, while the red asterisks represent the chief rays in the acquisition FOR. The $R_{CRMS}$ for the two FOV is shown in the figure, with the center of mass as the reference. The $R_{CRMS-S}$=0.0073 mm and the $R_{CRMS-A}$=0.1605 mm are indicated in the figure. The subscripts S and A represent the scientific FOV and the acquisition FOR, respectively. Our design results demonstrate that pupil aberrations and image plane aberrations are well corrected, surpassing our previous design results.

**Table 5. Detailed even-order aspheric coefficients**

| Surface | 4th-order term | 6th-order term | 8th-order term | 10th-order term | 12th-order term |
|---|---|---|---|---|---|
| SM | -1.36e-06 | 4.62e-10 | -1.72e-13 | 3.61e-17 | - |
| TM | 4.08e-03 | 1.06e-03 | -4.22-03 | -1.12e-03 | 4.75e-03 |
| QM | -5.87e-05 | 1.81e-06 | -4.15e-08 | 1.56e-10 | 2.04e-12 |

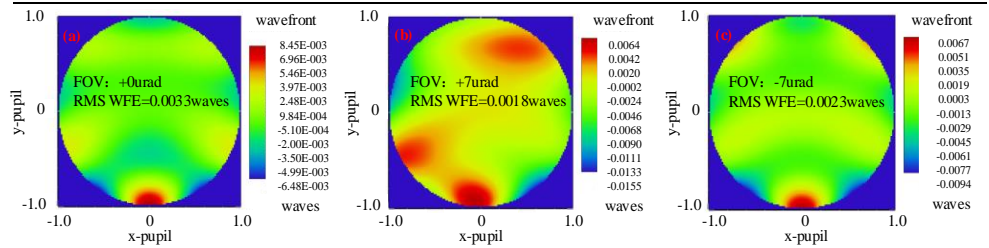

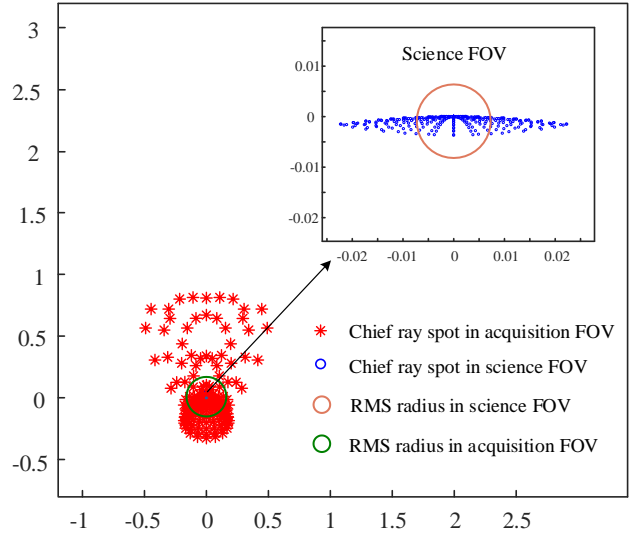

Fig. 7. Wavefront error over the scientific field of view

Fig. 8. Chief ray spot diagram in the exit pupil

We temporarily ignore the aberrations of other optical systems to investigate the TTL noise caused by the wavefront distortion of the telescope. The TTL coupling noise in the transceivers caused by the imperfect wavefront and continuous jitter during the GW detection can be obtained. The difference in the optical path length between the tilted measuring beam and Gaussian beam from the local optical bench can be calculated by evaluating the complex phase of the integral over the overlap term. We adopt the interference model and corresponding code provided by Y. Zhao et. al [43]. Thus, the following necessary quantities are defined in the similar way. The complex amplitude of the measuring beam on the detector can be expressed as $E_{\text{Flat}}(x, y, z)=\exp[-ik(x\sin\alpha+z\cos\alpha)+2\pi iW]$ when we set t=0. We fit the wavefront error in the measuring beam using the first 25 terms of the fringe Zernike polynomials, which can be expanded as:

$$W(\rho,\theta) = \sum a_i Z_i(\rho,\theta) \ (i = 4,9,16,25) + \sum A_j^{aber} Z_j(\rho,\theta) \quad (28)$$
$$(j = 5,7,10,12,14,17,19,21,23)$$

where $Z_n$ is Zernike fringe polynomial coefficients, $A_i$ is characterized by

$$A_i Z_i(\rho,\theta) = a_i Z_i(\rho,\theta) + a_{i+1} Z_{i+1}(\rho,\theta) \ (i = 5,7,10,12,14,17,19,21,23). \quad (29)$$

In Eq. (29), the cosine and sine terms of the Zernike aberrations are combined by the computation of trigonometric functions. The magnitude and orientation of the Zernike series expansion are listed in Table 6. The electric field of local reference fundamental Gaussian beam can be written as $E_{\text{Gauss}}(x, y, z)=\exp[-ik(x^2+y^2)/2q-ik(z-z_0)]$, where $z_0$ is the waist position. Then the phase information is calculated by complex amplitude [24]. The improved design specification meets the noise budget of 0.6nm/μrad anywhere. The maximum TTL over the entire ±300 μrad FOV is 0.0034nm/μrad. This performance is a significant improvement over previous version design.

Table 6. Magnitude and orientation of Zernike series expansion

| Mag/Ori | $a_4$ | $A_5/\theta_{56}$ | $A_7/\theta_{78}$ | $a_9$ | $A_{10}/\theta_{1011}$ | $A_{12}/\theta_{1213}$ |
|---|---|---|---|---|---|---|
| Waves/° | -1.45e-4 | 6.6e-3/-1.09 | 7.36e-4/1.56 | 9.2e-4 | 1.4e-3/-0.59 | 4.21e-4/1.55 |
| $A_{14}/\theta_{1415}$ | $a_{16}$ | $A_{17}/\theta_{1718}$ | $A_{19}/\theta_{1920}$ | $A_{21}/\theta_{2122}$ | $A_{23}/\theta_{2324}$ | $a_{25}$ |
| 1.4e-04/-1.32 | 6e-4 | 8.47e-4/0.37 | 1.5e-3/1.55 | -1.1e-3/-9e-3 | 5.41e-4/1.57 | 2.01e-4 |

## 7. Conclusion

In this paper, we systematically analyze the physical meaning of the pupil aberration and its impact on the optical path difference caused by the different fields of view in the telescope system for space GW observatories. This concept is successfully introduced into the design of the telescope which has strict requirements for the optical path noise caused by the field jitter. By exploring pupil aberration theory and applying the tools such as the chief ray spot diagram and $R_{CRMS}$, we established an automated correction algorithm for the pupil aberration and image wavefront aberration for the TianQin telescope system, achieving a high-performance optical design. Our research on pupil aberration theory provide valuable insights into the concept of ultra-low wavefront distortion, leading to a rapid convergence to TTL coupling noise. Furthermore, the image plane aberration maintains low wavefront distortion after the pupil aberration correction, with the wavefront aberration is less than $\lambda/200$ at the science FOV, well beyond the diffraction limit. Overall, our work demonstrates the critical role of correcting pupil aberrations in achieving optimal telescope performance. This approach to design has the potential to be significant in upcoming missions focused on detecting gravitational waves in space.

## Acknowledgments

This work was supported by the National Natural Science Foundation of China (12274156, 62205116); Science, Technology, and Innovation Commission of Shenzhen Municipality (JCYJ20190809100811375, JCYJ20210324115812035).

## Disclosures

The authors declare no conflicts of interest.